# Enabling Rapid Calibration of BCI Systems that Detect Movement-Related Cortical Potentials in Children with Cerebral Palsy[1]

R. Saadatyar, *Member, IEEE*, D. Damiano, and A. Behboodi, *Member, IEEE*

***Abstract*—Brain-computer interface neurofeedback (BCI-NFT) can be employed for neurorehabilitation by reinforcing task-contingent cortical activity and peripheral sensorimotor activity; however, its clinical adoption—particularly in pediatric populations—is limited by extensive session-specific calibration requirements. This study presents a deep learning framework for detecting motor intent using cortical potentials (MRCPs) in children with cerebral palsy (CP), with the explicit goal of minimizing calibration burden while maintaining high decoding accuracy. EEG was recorded from 16 sensorimotor channels during repeated ankle dorsiflexion tasks across 27 sessions in four children with CP. A bidirectional long short-term memory (Bi-LSTM) network was trained to discriminate motor intent from rest using 3-s pre-movement EEG epochs. The accuracies of seven training protocols were systematically evaluated, spanning within-session, cross-session, cumulative within-subject, and cross-subject learning, both with and without transfer learning. Results demonstrate that cumulative within-subject training without any in-session calibration and cross-subject pretraining followed by brief fine-tuning achieved the highest performance, with mean accuracies of 0.92 and 0.91, respectively. Both protocols outperformed conventional single-session and session-to-session approaches ($p < 0.01$), and their performance remained robust across calibration fractions and participants. Importantly, the cumulative within-subject protocol eliminated the need for explicit calibration once sufficient prior data were available, while the cross-subject transfer protocol enabled rapid personalization using minimal data. The proposed framework advances the clinical feasibility of BCI-NFT systems for pediatric neurorehabilitation.**

## I. INTRODUCTION

BRAIN–COMPUTER interface (BCI) technologies are increasingly used in neurorehabilitation to enhance motor function, particularly after non-progressive neurological injury such as stroke. By decoding brain activity in real time, BCI can deliver closed-loop, task-specific sensory feedback and motor assistance, i.e., neurofeedback, strengthening the sensorimotor loop [1], [2], [3]. Thereby, BCI neurofeedback training (BCI-NFT) can drive motor learning and neuroplasticity. BCI-NFT has the potential not only to assist and compensate for lost function but also to facilitate reorganization of impaired sensorimotor circuits [2], [3].

Two learning paradigms commonly reinforce BCI-NFT: operant conditioning and associative (classical) learning. In operant conditioning, users volitionally modulate a feature of their cortical activity, such as sensorimotor rhythms, during a motor task, often imagination of a movement, i.e., motor imagery (MI), to maximize reward-contingent feedback [4], [5]. MI-based control can be effective but may demand substantial mental training for some users [4], [5]. In associative learning paradigms, naturally occurring cortical features—such as movement-related cortical potentials (MRCPs) that emerge during an imagined (MI) or voluntary motor task, which is often an attempt to move, i.e., motor attempt (MA), are paired with precisely timed sensory feedback. This feedback is typically delivered through functional assistance or targeted neuromuscular stimulation, creating a contingent link between the individual's motor intent and the sensory feedback [6], [7], [8], [9]. Because it uses an intrinsically-occurring cortical feature, the associative learning paradigm does not require mastery and is intuitive, in contrast to operant conditioning. Clinically, recent systematic reviews indicate that BCI-NFT interventions can improve upper limb function in stroke rehabilitation, with generally stronger evidence in the subacute phase [10]. Upper limb functional improvement induced by BCI-driven functional electrical stimulation lasted 30 weeks post-intervention in a controlled trial by Biasiucci et al. [1]. Beyond stroke, applications of BCI-NFT in individuals with cerebral palsy (CP), the most commonly diagnosed child-onset motor disorder, remain limited.

To date, only two major BCI-NFT studies have been conducted in children with CP [11], [12]. Both targeted upper-limb training, relied on MI and operant conditioning, and reported positive motor outcomes. However, MA-based associative learning may be more intuitive and easier to perform

This research was supported by the Intramural Research Program of the National Institutes of Health (NIH). The contributions of the NIH author(s) are considered Works of the United States Government. The findings and conclusions presented in this paper are those of the author(s) and do not necessarily reflect the views of the NIH or the U.S. Department of Health and Human Services.

D. D. is with the Neurorehabilitation and Biomechanics Research Section, Rehabilitation Medicine Department, Clinical Center, National Institutes of Health, Bethesda, MD 20892 USA (e-mail: damianod@cc.nih.gov).

A. B. is with the Department of Biomechanics, University of Nebraska at Omaha, Omaha, NE 68182 USA, (e-mail: abehboodi@unomaha.edu

for pediatric populations, particularly for children with cognitive challenges. MA tasks are also more directly relevant to motor rehabilitation than MI tasks. Therefore, our overarching objective was to develop a BCI-NFT system for children with CP that integrates both associative learning and MA to enhance intuitiveness, engagement, and therapeutic relevance.

In BCI-NFT systems that employ an associative learning paradigm, the reliable detection of motor tasks (MI or MA) is essential for creating an effective association [9]. Traditional machine learning pipelines, for example, linear discriminant analysis (LDA) and support vector machines (SVM), have reported motor intent detection accuracies in the ~70–86% range depending on paradigm and settings across healthy and post-stroke cohorts [13], [14], [15]. However, movement detection in those with CP is characterized by noisy EEG, inconsistent motor attempts, and substantial session-to-session variability; thus, reliance on handcrafted features, an essential component of ML, and (typically) linear decision boundaries can be limiting for the nonlinear, time-variant, and nonstationary dynamics of EEG.

Deep-learning models, such as multilayer perceptrons and convolutional neural networks (CNNs), has the ability to learn latent features in EEG and, once trained, may require minimal pre-processing of EEG [16]. Deep Learning has increasingly demonstrated strong potential for EEG decoding because of its ability to extract discriminative spatiotemporal features with minimal to no signal conditioning, achieving detection accuracy of 70% to 80% [17].

A major practical limitation across existing EEG-based BCI systems, whether classical machine learning or deep learning, is the substantial calibration burden required at every therapy session. Calibration protocols in the BCI-NFT systems fall broadly into two categories: (1) systems requiring same-day calibration at each therapy session**,** and (2) systems relying on a single baseline calibration session for the entire intervention.

Most ML-based BCI systems that were used for neurorehabilitation fall into the first category, typically requiring 20–80 labeled trials per session before training can begin. For example, Chowdhury et al. [18], [19] required 80 calibration trials per session, achieving MI accuracies of ~70–75%. Osuagwu et al. recalibrated daily using 20 motor-attempt trials (70–80% accuracy) [20].

In MRCP-based BCIs Ibáñez et al. (~70–80% accuracy) [21] and Bhagat et al. (75–85% accuracy**)** [22] also relied on session-specific calibration. Associative-learning paradigms impose comparable requirements: Mrachacz-Kersting et al. required 30 self-paced motor attempts per session to determine individualized MRCP's peak negativity timing [6], [23].

Only a minority of systems rely on a single calibration session. Biasiucci et al. conducted one extended MRCP calibration session (90 motor attempts) and reused the machine learning classifier for all therapy days, achieving an offline accuracy of 80–85% [1]. More recently, Rao et al. also showed that calibration using 100 MI trials collected in a single session could support achieving stable classification performance (approximately 75–85%) across ten intervention sessions without requiring daily recalibration [24].

More recently, Jochumsen et al. systematically evaluated multiple calibration strategies for MRCP-based movement-intention detection in individuals with CP using machine learning classifiers, including temporal, spectral, and template-based EEG features combined with random forest classifiers [25]. Their highest performance was achieved using same-day, within-session calibration, yielding approximately 75% accuracy in participants with CP. However, performance declined to approximately 60% when models were trained on prior sessions from the same participant and dropped further to approximately 40–53% in cross-subject calibration scenarios. Importantly, the authors noted that conventional calibration procedures may require 15–30 minutes prior to therapy, reducing available training time and posing substantial challenges for children with limited attention span.

In deep-learning–based BCIs, calibration demands often become even more pronounced. In MI decoding, Roy et al. demonstrated that a hybrid CNN–LSTM architecture could achieve approximately 70% accuracy; however, this performance relied on training the model with hundreds of subject-specific MI trials collected across multiple sessions, and it was not evaluated if the accuracy can be transferred to upcoming sessions with a brief or no in-session calibration [26]. Similarly, Tibrewal et al. reported ~69% accuracy using a compact CNN trained on MI EEG from 54 participants, but the evaluation relied on within-subject training using large individualized datasets (on the order of ~100–300 trials per participant) [27]. More advanced temporal-convolutional architectures, such as EEG-TCNet, achieved 75–85% MI accuracy by modeling long-range temporal dependencies; nevertheless, these results were obtained using full-session training data comprising several hundred trials per subject, with no explicit strategy to minimize per-session calibration [28]. These studies evaluated performance exclusively in offline settings and were not evaluated in real-time operation.

Far fewer deep-learning approaches have been reported for MA–based decoding, particularly in clinical populations. Behboodi et al. demonstrated that an LSTM-based MRCP detector could identify motor attempts in a child with cerebral palsy (CP) with ~81% accuracy; however, the system initially required 50 calibration trials per session, later reduced to 20 trials using transfer learning. Beyond this single pediatric case report, deep-learning–based MRCP decoding has not yet been systematically applied in children with CP, highlighting a critical gap in the literature [29].

In deep-learning–based BCIs, transfer learning—the reuse of knowledge learned from prior subjects or sessions to initialize or adapt models for new users—has emerged as a promising strategy to mitigate calibration burden by addressing subject-to-subject and session-to-session variability in EEG signals. Foundational work by Jayaram et al. (2016) formalized transfer-learning paradigms for BCIs and demonstrated that sharing information across subjects or sessions can substantially reduce the number of subject-specific calibration trials required to achieve stable decoding performance [30]. Building on this framework, several deep-learning approaches have explored cross-subject and calibration-efficient adaptation primarily in motor-imagery (MI) paradigms. For example, Wang et al. (2025) proposed a task-free transfer-learning strategy that adapts MI classifiers using resting-state EEG without requiring task-labeled calibration data from new users s [31], while Chowdhury et al. (2023) demonstrated improved cross-subject MI decoding using a multi-branch convolutional neural network [32]. Similarly, Lu et al. (2023) showed that hybrid domain-adaptation with few-shot fine-tuning can substantially improve cross-subject generalization in EEG-based emotion recognition [33]. Collectively, these MI- and affective-EEG–based studies report classification accuracies typically in the ~65–85% range while reducing reliance on extensive subject-specific retraining, highlighting growing interest in calibration-efficient decoding.

Transfer learning has only recently been applied explicitly to reduce calibration in rehabilitation-oriented BCIs, where transfer learning aims to minimize or eliminate repeated session-specific recalibration while preserving stable motor-intention detection. Fahimi et al. (2018) demonstrated that an end-to-end deep convolutional neural network pretrained across subjects could be transferred to new users using a reduced calibration set of approximately 20–40 MI trials, while maintaining decoding accuracies in the ~70–80% range, comparable to fully trained subject-specific models [34], while Kueper et al. (2024) investigated a cross-task transfer-learning strategy for exoskeleton-assisted rehabilitation in which classifiers trained on EEG recorded during bilateral arm movements were transferred to detect unilateral movement intentions. Their approach achieved balanced accuracies of approximately 75–85% while eliminating the need for dedicated calibration sessions, suggesting that transfer learning may reduce setup burden in rehabilitation-oriented BCIs [35].

Despite these advances, in rehabilitation BCI, transfer learning has been evaluated almost exclusively on motor-imagery paradigms in adult populations and does not address the unique challenges of MRCP-based MA detection, which relies on low-frequency, non-oscillatory EEG features. Moreover, these strategies have not been validated in pediatric or neurologically impaired populations, where EEG variability, attention fluctuations, and fatigue effects are amplified. In such settings, systems requiring same-day calibration reduce available therapy time and increase cognitive and physical burden, particularly for children. Together, these limitations underscore the need for calibration-efficient, high-accuracy MRCP decoding frameworks tailored to pediatric neurorehabilitation a gap directly addressed by the deep-learning and transfer-learning approach proposed in this study.

The main objective of this study was to develop a deep-learning model capable of detecting motor intent during MA, which is referred to as *motor intent*, in children with CP from MRCPs, utilizing transfer learning to achieve high detection accuracy with minimal calibration time.

To this end, we first developed a deep-learning model, then evaluated the model's motor intent detection with and without transfer learning. We then compared how accuracy changed when applying three different transfer-learning strategies: (1) session-to-session transfer, (2) within-subject transfer, and (3) between-subject transfer. We hypothesize that the accuracy of our deep learning model with between-subject transfer learning provides the highest accuracy with minimal calibration. This work aims to advance the clinical feasibility and scalability of BCI-NFT, particularly for pediatric neurorehabilitation.

## II. Methods

### A. Participants

The study analyzed previously collected data from four individuals with cerebral palsy (mean age = 16.4 years; 3 female /1 male). Participants were recruited at the NIH Clinical Center to evaluate our previously reported BCI-NFT platform with adaptations for pediatric cohorts [29]. The experimental paradigm and software–hardware stack followed our prior reports, with the present work focusing on MRCP-based detection during ankle dorsiflexion [16], [29]. The protocol was approved by the NIH Institutional Review Board (Protocol #13-CC-0110). Parent consent and child assent were obtained. Participants' details can be found in Table I.

TABLE I.
Participants' Details.

| ID | Age (years) | Sex | GMFCS |
|---|---|---|---|
| CP01 | 11.50 | M | I |
| CP02 | 19.49 | F | III |
| CP03 | 18.17 | F | II |
| CP04 | 16.42 | F | II-III |

### B. Experimental Protocols

During each BCI-NFT session, participants sat in a long sitting posture facing a monitor. Two inertial measurement units (IMUs) were placed on the midfoot and lateral shank of the paretic limb to track ankle kinematics. During each session (i.e., each training day), data collection was organized into blocks of 5 trials, with each trial lasting 10 s and consisting of three phases. Each trial began with a 3 s relaxation phase, followed by a 3 s *preparation phase* during which a visual loading bar progressively filled on the screen. Once the bar was fully loaded, participants were cued at the 6 s mark to perform ankle dorsiflexion (DF) as quickly and forcefully as possible and maintain the contraction for 4 s (motor execution phase). Additional details regarding the experimental protocol can be found in our previous publication [29].

Session and trial counts per participant were as follows: CP01, 10 sessions (97-135 trials per session); CP02, 5 sessions (56-85); CP03, 6 sessions (75-100); and CP04, 6 sessions (60-120). In total, the dataset comprised 27 sessions and 2,637 trials.

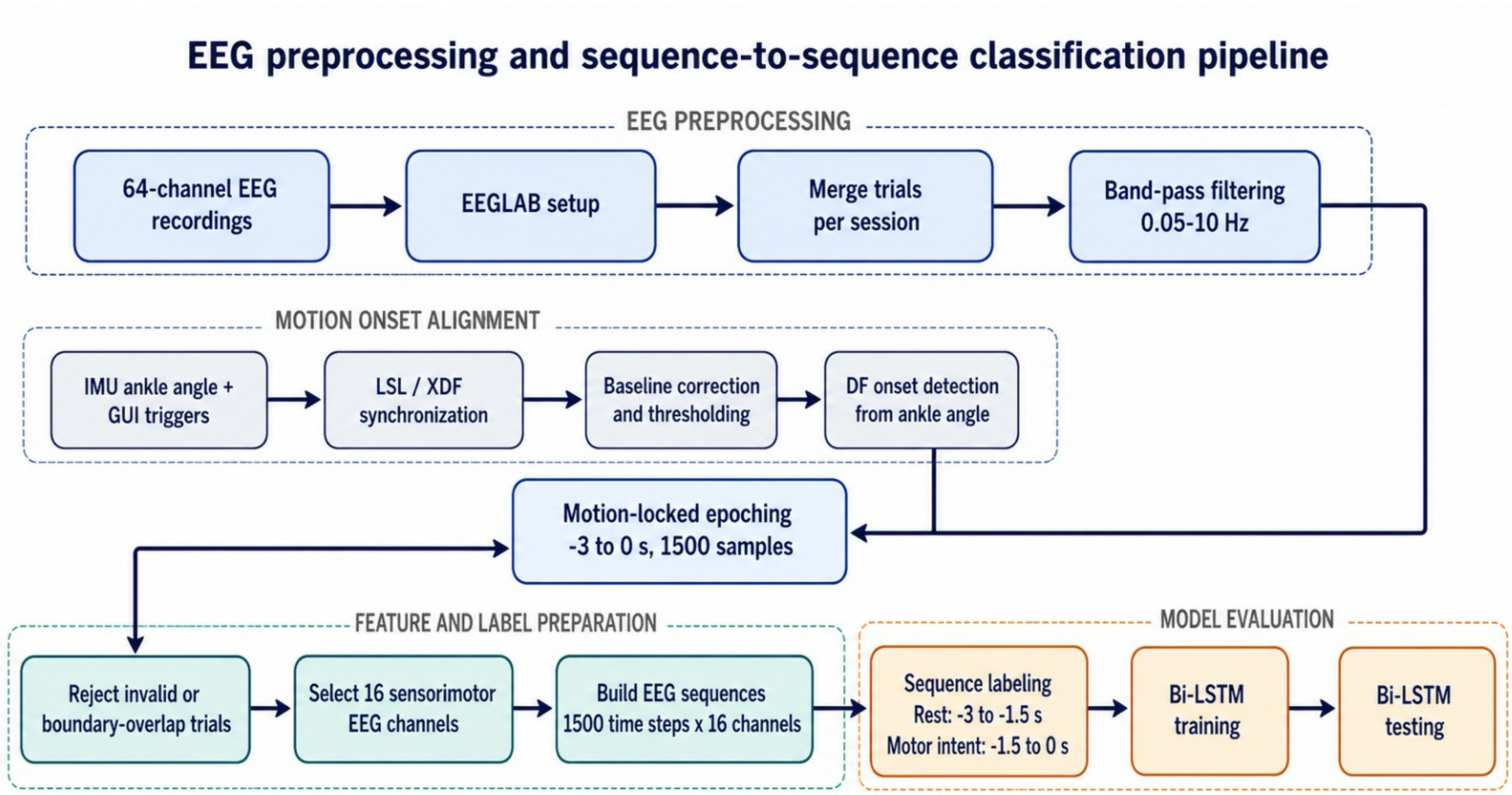

Fig. 1. The classification pipeline includes EEG preprocessing, ankle dorsiflexion detection, feature preparation, and model evaluation.

A graphical user interface (GUI), developed in Unity (Unity Software Inc., San Francisco, CA), displayed on-screen cues and instructions and provided real-time visual feedback of ankle angle [29]. The GUI generated event markers, which we call triggers, corresponding to each trial phase—relaxation, preparation, dorsiflexion onset, and hold—which were streamed to a custom MATLAB (MathWorks, Natick, MA) script via the Lab Streaming Layer (LSL). These markers were used to time-stamp the EEG and ankle-angle signals for precise alignment with task events [29].

### C. Data Acquisition

EEG was recorded from 64 electrodes (actiCap/Brain Products) at 500 Hz with FPz as ground and FCz as reference, arranged according to the international 10–20 layout [29]. Conductive gel was applied until electrode–scalp impedance was below 20 kΩ, as indicated by the green LED status on each electrode [29]. EEG and kinematic streams were synchronized via event markers sent from the GUI through LSL, the Lab Streaming Layer [29].

### D. EEG Pre-processing

Pre-processing was performed using the EEGLAB toolbox [36] in the MATLAB script. At the first stage, all of the trials from each session were merged. Signals were then band-pass filtered at 0.05–10 Hz using a zero-phase second-order Butterworth filter with *filtfilt* to retain slow cortical components characteristic of MRCPs while reducing higher-frequency noise. The MRCP frequency is based on the range reported in Mrachacz-Kersting et al. [6], [23].

### E. Epoching the EEG data:

To ensure precise time-locking to the actual physical movement rather than relying solely on GUI-generated trigger events, DF onset was detected directly from ankle angle and defined as time zero. Ankle angles were acquired from IMUs streamed to the Unity-based GUI and synchronized with EEG and trigger events through LSL recordings stored in XDF format.

DF onset was identified on a trial-by-trial basis using a standard deviation thresholding algorithm applied to the ankle angle signal. First, a 1 s baseline period preceding the dorsiflexion cue was identified using the GUI-generated event markers. Signal drift or baseline bias within this interval was then estimated and subtracted from the full angle trajectory. The mean and standard deviation of the baseline segment were subsequently computed, and movement onset was defined as the first time point exceeding a threshold equal to the baseline mean plus three standard deviations (mean + 3SD).

EEG data were then epoched into 3 s windows referenced to the detected DF onset (time zero), extending from −3 s to movement onset (0 s), resulting in 1,500 samples per epoch at a sampling rate of 500 Hz (Figure 1). Trials lacking a valid kinematic onset marker or containing merged epoch boundaries were automatically excluded from analysis. These 3 s epochs were subsequently used for MRCP-based motor-attempt detection.

### F. Data Labeling for Sequence-to-Sequence Analysis

In healthy individuals, the intention to move (*motor intent*) generates a motor command in the brain, which subsequently leads to muscle activation and visible movement. This motor command is reflected in the MRCP—a slow, low-frequency

Fig. 2. Detection model architecture. The network processes sequences bidirectionally, updating cell and hidden states through input/forget/output gates. B) Prior to training, the data is labeled into two classes: Rest and Motor intent, with reference to the dorsiflexion (DF) onset. C) LSTM memory unit.

EEG deflection characterized by a gradual negative shift beginning approximately 1.5–2 s before movement onset. This early negative trend enables the detection of motor intent prior to muscle activation. The most prominent part of the MRCP is the peak negativity, which typically occurs within ~500 ms of movement onset, as described in a recent review [14].

To detect the MRCP feature, we utilized the sensorimotor subset (16 channels) of our 64-channel EEG, where MRCP is typically present: {FC5, FC3, FC1, FCz, FC2, FC4, C3, C1, Cz, C2, C4, CP3, CP1, CPz, CP2, CP4}. Then, we trained a deep-learning-based classifier, which was trained to solve a two-class classification problem (rest vs. *motor intent*).

For processing these signals, we employed a deep Bidirectional LSTM (Bi-LSTM) network operating in a sequence-to-sequence manner.

To maximize the likelihood of capturing the MRCP deflection while providing perfectly balanced training data, each 3-s continuous trial (1,500 samples) was temporally partitioned into two contiguous, equally-sized blocks of 750 samples. Relative to the exact physical dorsiflexion onset (time zero), the sequential labeling scheme was formulated as follows:

- **Rest** (Class 0): −3000 ms to −1500 ms (first 750 consecutive samples).
- **Motor intent** (Class 1): −1500 ms to 0 ms (subsequent 750 consecutive samples).

This contiguous temporal partitioning enables the Bi-LSTM model to not only isolate the most prominent window wherein MRCPs occur but also continuously learn the dynamic time-series transition from an idle rest state to active motor preparation.

## G. Deep Learning

*Long Short-Term Memory (LSTM)*

EEG signals are inherently temporal and exhibit dependencies that span many time steps, particularly in MRCPs, which evolve gradually over hundreds of milliseconds prior to movement onset. Standard recurrent neural networks (RNNs) often struggle to model such long-range temporal relationships due to the vanishing gradient problem, wherein error gradients decay during backpropagation and limit learning from earlier time points. Long short-term memory (LSTM) networks address this limitation by introducing gated memory cells that selectively retain or discard information over time, enabling stable gradient flow and effective modeling of extended temporal structure. As a result, LSTMs are well-suited for EEG classification tasks that require integration of slow, low-frequency neural dynamics across time [37].

An LSTM unit maintains a cell state $c_t$ and a hidden state $h_t$. Information flow is controlled by forget, input, and output gates (sigmoid activations). For the time step $t$, given input $x_t$, previous hidden state $h_{t-1}$, and the previous cell state $c_{t-1}$:

- Forget gate

$$f_t = \sigma(w_f.[h_{t-1}, x_t] + b_f \quad (1)$$

- Input gate and candidate values

$$i_t = \sigma(w_i[h_{t-1}, x_t] + b_i \quad (2)$$

$$C_t^{\infty} = tanh(w_c[h_{t-1}, x_t] + b_c \quad (3)$$

Cell state update

$$C_t = f_t \odot C_{t-1} + i_t \odot C_t^{\infty} \quad (4)$$

- Output gate and hidden state

$$o_t = \sigma(w_o[h_{t-1}, x_t] + b_o \quad (5)$$

$$h_t = o_t \odot tanh(C_t) \quad (6)$$

Here, σ denotes the logistic sigmoid, *tanh* the hyperbolic tangent, and ⊙ element-wise multiplication. The combination of these mechanisms enables LSTMs to preserve long-term dependencies while remaining adaptable to new input data.

*Bi-Directional Long Short-Term Memory (Bi-LSTM)*

While LSTMs capture temporal dependencies, a standard LSTM processes the sequence in the forward direction only, so the hidden state at time *t* depends exclusively on past inputs. In EEG analysis, it is often beneficial to exploit context on both sides of an event, e.g., MRCP patterns before and after peak negativity. A Bi-LSTM extends the LSTM with two parallel layers: one scans forward and the other backward; their hidden states are concatenated to form the representation at each time step:

$$h_t^{bi} = \left[h_t^{forward}; h_t^{backward}\right] \quad (7)$$

In our setting, the Bi-LSTM operates on 3s MRCP segments prior to the cue to DF event. Each sequence has a length *T* = 1500 at 500 Hz, and each time-step input is a 16-channel EEG vector $x_t \in R^{2}d$. The structure of the Bi-LSTM network and the internal operations of an LSTM cell, including the input, forget, and output gates, are illustrated in Figure 2.

*Network Architecture and Training Setup*

The Bi-LSTM model assigns a class label (rest or motor intent) to each step within a 3s EEG epoch. Sequences correspond to the 3 s MRCP windows with 16 sensorimotor channels described in Section II-D.

To efficiently handle long sequences (1500 time-steps) and extract robust spatio-temporal features, our network architecture was significantly extended. The network begins with a 1D Batch Normalization layer for channel-wise dynamic standardization of EEG voltage ranges, followed by a temporal Max-Pooling layer (pool size = 5) to compress the sequence length and reduce computational overload while expanding the temporal receptive field. The core temporal representation is built upon two stacked Bi-LSTM layers comprising 128 and 64 hidden units per direction, respectively. To mitigate overfitting and prevent the network from relying on single dominant EEG channels, a Spatial Dropout layer (dropout rate = 0.5) is inserted between the LSTM layers, followed by standard Dropout. As in (7), the forward and backward hidden states at each time step are concatenated to form a unified representation that captures temporal context from both past and future samples $h_t^{bi}$. Finally, the compressed temporal representation is up-sampled to its original sequence length, Frame-wise class probabilities are computed by:

$$y_t = Softmax\left(w_y . h_t^{bi} + b_y\right),$$
$$W_y \in R^{Kx2H}, b_y \in R^K, K = 2 \quad (8)$$

Training used the Adam optimizer configured with an initial learning rate of 0.001 and an L2 weight decay penalty of 1e-3 to prevent catastrophic weight growth. To imbalances inherent in continuous block processing, a class-weighted Negative Log-Likelihood Loss (analytically equivalent to weighted categorical cross-entropy) was utilized. The mini-batch size was set to 10% of the number of training sequences (minimum 1), and the number of epochs was adjusted automatically based on the training set size $N_{seq}$.

The sequence length was set to the maximum available window to preserve the full temporal evolution of the EEG signal, which is particularly important for capturing the slow pre-movement dynamics of MRCPs that unfold over hundreds of milliseconds to seconds. During training, global L2-norm gradient clipping was applied with a threshold of 1, where the L2 norm—defined as the Euclidean magnitude of the gradient vector across all trainable parameters—was constrained to prevent excessively large weight updates. This procedure mitigates the exploding gradient problem during backpropagation through long temporal sequences, thereby ensuring stable model convergence. Table II summarizes the configuration for training the model with and without transfer learning.

Performance is reported as accuracy, defined per epoch as the mean proportion of correctly classified time steps, $\hat{y} = y_t$, across its *T* samples:

$$Accuracy = \frac{1}{T}\sum_{t=1}^{T}[\hat{y} = y_t] \quad (9)$$

where $\hat{y}$ is the model prediction, and $y_t$ is the labeled sequence. In addition to accuracy, the F1-score, defined as the harmonic mean of precision and recall, was computed to evaluate how well the model balanced correctly detecting motor intent (true positives) while minimizing false-positive detections. This metric was particularly important given the need for reliable associative timing in BCI paradigms.

To further evaluate the discriminative performance of the proposed Bi-LSTM classifier across the seven training protocols, receiver operating characteristic (ROC) analysis was performed using the pooled predictions from all participants. ROC curves were generated by varying the decision threshold applied to the model output probabilities and plotting the true positive rate (TPR; sensitivity) against the false positive rate (FPR; 1 − specificity). The area under the ROC curve (AUC) was computed as a threshold-independent measure of classification performance, where an AUC of 0.5 indicates chance-level discrimination and an AUC of 1.0 indicates perfect classification. ROC analysis was performed separately for each protocol using the held-out test data from all sessions and participants combined

*Transferring the Model*

To transfer a trained model, we first froze the entire pretrained feature extraction pipeline, encompassing the stacked Bi-LSTM layers, preventing their weights from being updated. We then fine-tuned only the dense classification head (the fully connected output layer) using the calibration subset from the current session (session n). This approach preserves the generalized MRCP representation learned from prior data while allowing the classification boundary to adapt to session-specific EEG characteristics.

### *H. Training and Transfer-Learning Evaluation Protocols*

The goal of this analysis was to identify the optimal training and transfer-learning protocol that minimizes calibration requirements while maintaining high detection accuracy. To this end, we evaluated seven protocols (Figure 3) encompassing within-session, cross-session, and cross-subject training schemes, both with and without transfer learning. In this study, high accuracy is defined as ≥85%, corresponding to the upper range of accuracies reported in prior studies reviewed in the Introduction.

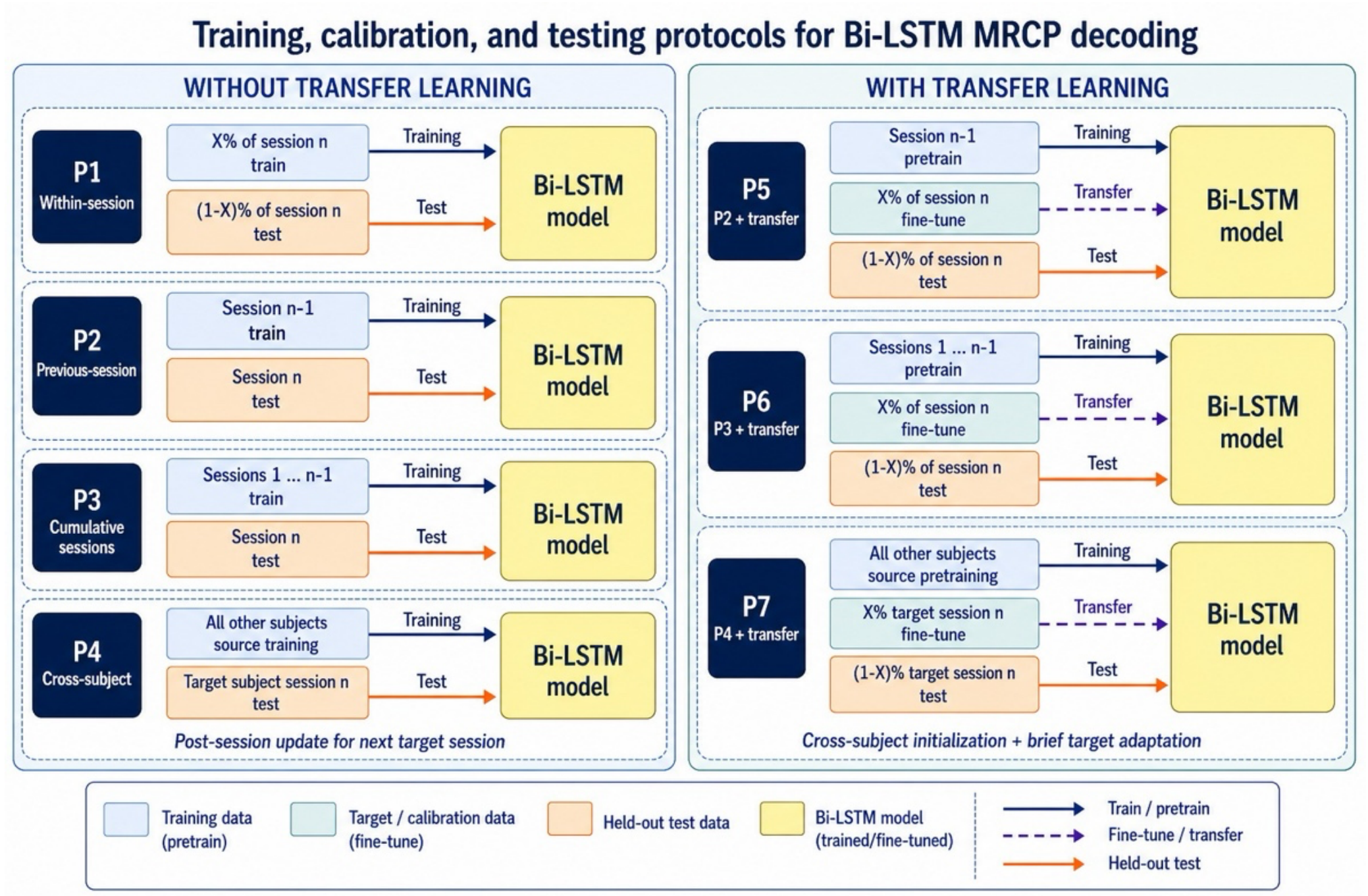

Fig. 3. Training and testing strategies for the Bi-LSTM model with and without transfer learning.

All reported results represent the mean performance across five-fold cross-validation evaluations (k = 5). When applicable, a fixed calibration proportion $X$(e.g., 25% or 50%) of the target session was incorporated during model development. In this study, calibration refers to the use of a subset of data from the session designated for testing the model (target session) either to train the model from scratch or to fine-tune a pretrained model prior to testing the model on the remaining unseen portion of the target session data.

*Protocols without Transfer Learning*

Four protocols were implemented that used no data for calibration and did not incorporate transfer learning:

- Protocol 1, within-session (train and test on the current session): The model was trained on $X\%$ of EEG data from session $n$ (the current session) utilizing k-fold cross validation and tested on the remaining $(1 - X)\%$. This protocol establishes single-session performance without leveraging prior sessions. This protocol was considered our base model.
- Protocol 2, previous session (train with previous session → test on current session): The model was trained on session $n$-$1$ and tested directly on session $n$ of the target participant. This evaluates cross-session generalization without within-session calibration.
- Protocol 3, culminative sessions (cumulative train on previous sessions→ test on current): The model was trained on all available earlier sessions (sessions 1 to n − 1) from the target participant and tested on session $n$. This examines whether aggregating multiple prior sessions of the same subject improves generalization in the absence of calibration.
- Protocol 4, cross-subject: A generalized base model was first pretrained using all sessions from all participants except the target participant (source subjects). The pretrained model was then evaluated sequentially on the target participant's sessions, thereby providing a true zero-calibration cross-subject baseline for the initial session. Although no target-subject calibration data were used prior to the first evaluation, the model was incrementally updated after each session. Specifically, following evaluation on a given target session, that session's data were incorporated into the training dataset, and the model was retrained prior to evaluation on the subsequent target session.

*Protocols with Transfer Learning*

In the protocols with transfer learning, the pretrained model was calibrated (transferred) on a small calibration segment ($X\%$) of session $n$, after which performance was evaluated on the remaining $(1 - X)\%$ of that session. This approach aims to improve session-to-session and subject-to-subject generalizability and adaptation of the model.

- **The original Protocol**, Protocol 5 (Protocol 2 + Transfer Learning): The model was trained on the target participant's session $n - 1$, then calibrated (transferred) to the target session using $X\%$ of the session's data and tested on the remaining $(1 - X)\%$ of the EEG data. This tests whether minimal calibration can compensate for cross-session variability. Note that this protocol was the original model that was used in our developed BCI-NFT system [29].

- Protocol 6 (Protocol 3 + Transfer Learning): The model was pretrained on all earlier sessions (sessions 1 to n − 1), then calibrated using *X%* of session *n*, and tested on the remainder. This protocol evaluates whether cumulative pretraining combined with limited calibration enhances robustness.
- Protocol 7 (Protocol 4 + Transfer Learning): This protocol is identical to Protocol 4 but adds transfer learning. The cross-subject pretrained model was calibrated on *X%* of session *n* and evaluated on the remaining *(1 − X)%*. This assesses whether combining broad cross-subject pretraining with a small amount of subject-specific calibration yields strong generalization.

### I. Statistical Analysis

A One-Way Analysis of Variance (ANOVA) was performed at the session level using the protocol as the within-subject factor with seven levels. The significance threshold (α) was set at 0.05 for individual participants. To rigorously assess the pairwise differences between specific protocols following the overarching ANOVA, post-hoc paired comparisons were conducted utilizing Tukey's Honestly Significant Difference (Tukey HSD) test.

TABLE II.
Configuration of the Bi-LSTM model with and without transfer learning.

| Parameter | With Transfer |
|---|---|
| Number of Layers | Eight: BatchNorm → MaxPool → 2×Bi-LSTM → UpSample → FC → Softmax (Figure 2.A) |
| Optimizer | Adam (LR=0.001, Weight Decay=1e-3) |
| Loss | Class-weighted Categorical Cross Entropy |
| Maximum epochs | *With **transfer***: 2×[$N_{seq}$]; *Without **transfer*** [$N_{seq}$] |
| Mini-Batch Size | [$N_{seq}$/10] (min 1) |
| Number of Classes | 2 |
| Regularization | Spatial Dropout (0.5), Standard Dropout (0.5), Gradient Clipping (max_norm=1) |
| Bi-LSTM hidden units | Layer 1: 128 / dir, Layer 2: 64 / dir |
| Learning Rate Factors | All layers learnable; *to **Transfer*** Feature Extractor Frozen; Classifier Head (FC Layer) learnable |
| Output mode | Sequence (frame-wise hidden states) |

$N_{seq}$ = number of training sequences in the run; dir=direction; X ∈ {25%, 50%}; global - $\ell_2$

## III. Result

### A. Overall Performance

Across four participants (CP01–CP04), the Bi-LSTM was evaluated under seven training protocols comprising four without transfer learning (Protocols 1–4) and three with transfer learning (Protocols 5–7). Figure 4A presents the mean accuracy for each protocol averaged across participants, whereas Figure 4B shows the individual accuracies for CP01–CP04 across all protocols, highlighting the degree of between-participant variability. In all of the reported accuracies, where applicable, 25% of the target session's (session n) EEG data was used for calibration (X = 25%), evaluating performance on the strictly held-out 75% test split.

As depicted in Figure 4A, Protocols 7 and 4 achieved the highest overall accuracy and F1-scores, reaching 0.98 and 0.95, respectively (Fig. 4 and Table III). In contrast, Protocol 1, which strictly restricted training to a limited 25% within-session split without leveraging any historical data, yielded the lowest average accuracy (0.68).

### B. Subject-Specific Results

Across all four participants (CP01–CP04), the Bi-LSTM classifier showed a consistent pattern across evaluation protocols. Cumulative cross-subject pretraining (Protocol 4) and cross-subject pretraining coupled with brief calibration (Protocol 7) produced the strongest accuracies, ranging remarkably from 0.80 to 0.99 (Table II), and significantly outperformed all other protocols. Furthermore, cumulative within-subject training (Protocol 3) produced robust accuracies (averaging 0.86), while previous-session training (Protocol 2) yielded moderate performance (0.71–0.84 across subjects).

Through the lens of limited data, within-session training (Protocol 1, X=25%) consistently yielded the lowest baseline accuracy (averaging 0.68), primarily because the deep network structure lacked sufficient temporal data to optimally converge without historical transfer. Transfer learning protocols 5 and 6 demonstrated substantial improvements over their non-transfer counterparts, with Protocol 6 achieving strong performance (average 0.85 for X=25%) but still failing to match the near-perfect invariance of Protocols 4 and 7. Both CP02 and CP03 showed the weakest initial baseline performance in Protocol 1 (0.64 and 0.63, respectively), likely reflecting more variable MRCP morphology; however, their accuracy improved to ≥0.97 under Protocol 7. This demonstrates that relying uniformly on broad cross-subject data (Protocol 4) or combining it with calibrated transfer (Protocol 7) effectively overcomes individual morphological variabilities.

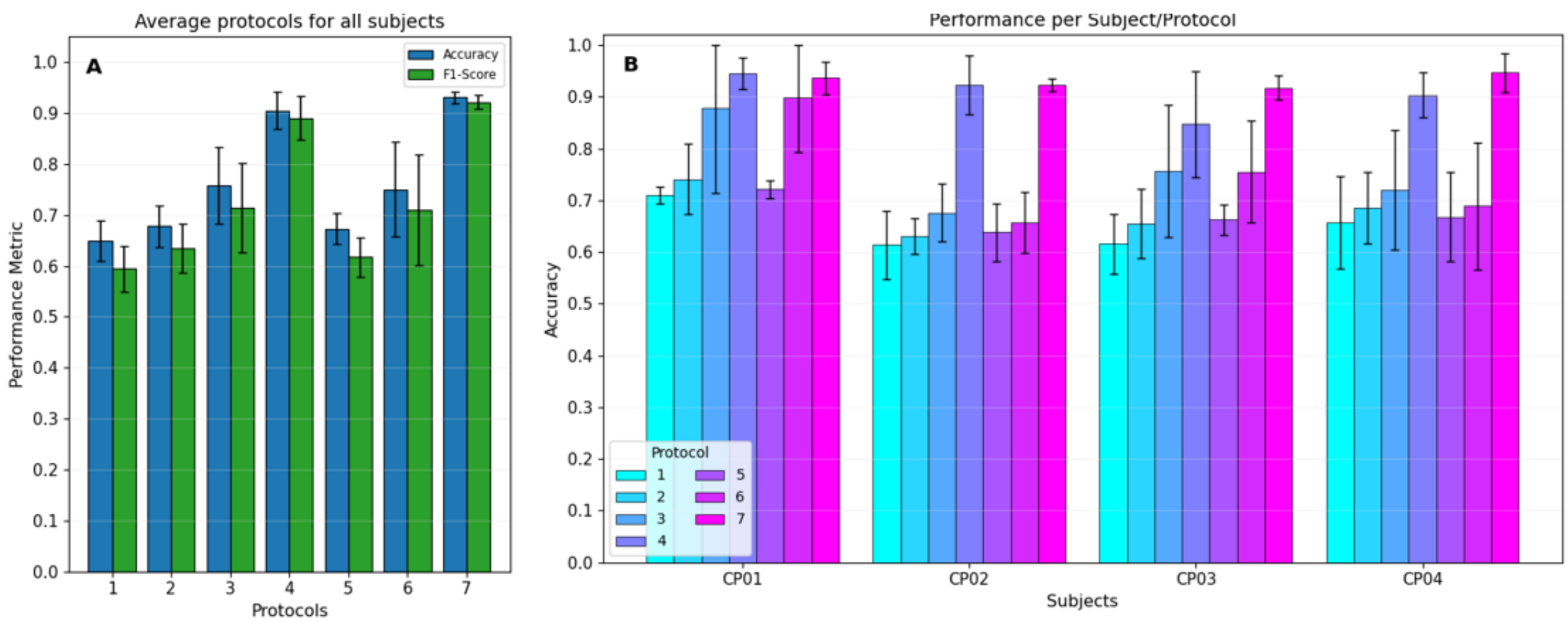

Fig 4. Our Bi-LSTM accuracy across seven protocols. (A) Mean accuracy and F1-Score over sessions, averaged across CP01–CP04. (B) Accuracy by participant (CP01–CP04) for all protocols. Error bars show standard deviation. Results are for *X*=25%; 25% of trials for training, 75% for testing.

### *C. Session by Session Accuracy (Table III)*

For the *X* = 50% test split, i.e., 50% within-session data used for training (Protocol 1) or transfer/calibration (Protocols 5, 6, 7)—accuracies were typically higher than the corresponding X=25% results due to the larger training/calibration portion. For instance, Protocol 1's average improved from 0.68 to 0.76. However, the overall ranking of the protocols remained strictly unchanged: Protocols 4 and 7 remained the best, while Protocol 1 (X=25%) remained the worst. Among transfer protocols, Protocol 6, which used multiple subject-specific sessions for pretraining the model, consistently outperformed Protocol 5, which only used the immediately preceding session for pretraining prior to transferring.

In Protocols 4 and 7, the model was initially trained using data from all other subjects and evaluated on Session 1 of the target subject. The target session data was then incrementally appended, and the model was subsequently evaluated on the next sequential session. In the case of Protocol 7, this step involved active transfer learning (fine-tuning the classification head) prior to testing. Notably, in both protocols, detection accuracy remained largely stable and exceptionally high across sessions (Table III), indicating that the initial robust pretraining adequately captured universal MRCP features, such that the sequential incorporation of additional uncalibrated subject-specific sessions did not cause catastrophic shifting or performance degradation.

TABLE III

Comparative Bi-LSTM performance across four subjects under non-transfer and transfer learning protocols. *X* indicates the percentage of the EEG dataset that was used for training (in case of Protocol 1) or calibrating (in case of Protocols,5,6,7) the model.

| Subjects | Accuracy without Transfer Learning | | | | | Accuracy with Transfer Learning | | | | | |
|---|---|---|---|---|---|---|---|---|---|---|---|
| | Protocol 1 | | Protocol 2 | Protocol 3 | Protocol 4 | Protocol 5 | | Protocol 6 | | Protocol 7 | |
| | X= 50% | X= 25% | | | | X= 50% | X= 25% | X= 50% | X= 25% | X=50% | X=25% |
| CP01 | 0.73±0.03 | 0.71±0.02 | 0.74±0.07 | 0.88±0.16 | 0.95±0.03 | 0.73±0.02 | 0.72±0.02 | 0.91±0.10 | 0.90±0.11 | 0.96±0.03 | 0.94±0.03 |
| CP02 | 0.69±0.04 | 0.61±0.06 | 0.63±0.03 | 0.68±0.06 | 0.92±0.06 | 0.68±0.04 | 0.64±0.05 | 0.70±0.05 | 0.66±0.06 | 0.91±0.04 | 0.92±0.01 |
| CP03 | 0.65±0.03 | 0.62±0.06 | 0.65±0.07 | 0.76±0.13 | 0.85±0.10 | 0.68±0.02 | 0.66±0.03 | 0.76±0.05 | 0.76±0.10 | 0.92±0.04 | 0.92±0.02 |
| CP04 | 0.74±0.03 | 0.66±0.09 | 0.68±0.07 | 0.72±0.17 | 0.90±0.04 | 0.74±0.02 | 0.67±0.09 | 0.80±0.08 | 0.69±0.12 | 0.95±0.03 | 0.95±0.04 |
| Avg | 0.70±0.04 | 0.65±0.04 | 0.68±0.04 | 0.76±0.07 | 0.91±0.04 | 0.71±0.03 | 0.67±0.03 | 0.79±0.08 | 0.75±0.09 | 0.94±0.02 | 0.93±0.01 |

TABLE IV

The Bi-LSTM accuracy performance under four protocols with non-transfer learning and three protocols without transfer learning for each subject. X indicates the percentage of the EEG dataset that was used for training (in case of Protocol 1) or calibrating (in case of Protocols,5,6,7) the model. The accuracy could not be calculated for the first session (S01) due to the nature of protocols 2, 3, 5, and 6.

| | Without Transfer Learning | | | | | With Transfer Learning | | | | | |
|---|---|---|---|---|---|---|---|---|---|---|---|
| | Protocol 1 | | Protocol 2 | Protocol 3 | Protocol 4 | Protocol 5 | | Protocol 6 | | Protocol 7 | |
| CP01 | X= 50% | X= 25% | | | | X= 50% | X= 25% | X= 50% | X= 25% | X= 50% | X=25% |
| S01 | 0.67±0.02 | 0.72±0.02 | ----- | ----- | 0.89±0.20 | ----- | ----- | ----- | ----- | 0.90±0.2 | 0.99±0.01 |
| S02 | 0.70±0.03 | 0.68±0.06 | 0.72±0.05 | 0.64±0.03 | 0.99±0.00 | 0.71±0.04 | 0.72±0.02 | 0.71±0.01 | 0.71±0.03 | 0.97±0.04 | 0.90±0.17 |
| S03 | 0.71±0.02 | 0.69±0.01 | 0.55±0.01 | 0.53±0.01 | 0.91±0.17 | 0.71±0.02 | 0.68±0.03 | 0.77±0.10 | 0.79±0.15 | 0.97±0.04 | 0.94±0.09 |
| S04 | 0.79±0.04 | 0.74±0.03 | 0.80±0.01 | 0.74±0.01 | 0.95±0.07 | 0.75±0.03 | 0.73±0.03 | 0.80±0.01 | 0.73±0.02 | 0.94±0.09 | 0.92±0.10 |
| S05 | 0.75±0.02 | 0.72±0.04 | 0.76±0.01 | 0.99±0.00 | 0.98±0.02 | 0.75±0.02 | 0.71±0.03 | 0.90±0.11 | 0.94±0.10 | 0.99±0.02 | 0.90±0.12 |
| S06 | 0.72±0.06 | 0.72±0.03 | 0.74±0.01 | 0.99±0.00 | 0.92±0.13 | 0.74±0.01 | 0.72±0.01 | 0.98±0.01 | 0.92±0.12 | 0.96±0.04 | 0.89±0.13 |
| S07 | 0.74±0.02 | 0.71±0.02 | 0.76±0.01 | 0.99±0.00 | 0.93±0.12 | 0.72±0.03 | 0.71±0.03 | 0.98±0.01 | 0.98±0.01 | 0.99±0.01 | 0.92±0.01 |
| S08 | 0.73±0.03 | 0.73±0.01 | 0.78±0.01 | 0.98±0.01 | 0.94±0.09 | 0.72±0.01 | 0.73±0.02 | 0.98±0.01 | 0.99±0.01 | 0.99±0.01 | 0.93±0.09 |
| S09 | 0.73±0.03 | 0.69±0.04 | 0.76±0.02 | 0.99±0.08 | 0.94±0.09 | 0.71±0.03 | 0.73±0.02 | 0.99±0.01 | 0.99±0.01 | 0.99±0.01 | 0.97±0.05 |
| S10 | 0.76±0.02 | 0.72±0.04 | 0.78±0.03 | 0.99±0.01 | 0.96±0.06 | 0.75±0.03 | 0.75±0.02 | 0.99±0.00 | 0.99±0.01 | 0.98±0.04 | 0.96±0.05 |
| S11 | 0.74±0.02 | 0.70±0.05 | 0.77±0.03 | 0.93±0.11 | 0.97±0.02 | 0.73±0.04 | 0.73±0.04 | 0.97±0.03 | 0.96±0.06 | 0.91±0.17 | 0.98±0.02 |
| Avg | 0.73±0.03 | 0.71±0.02 | 0.74±0.07 | 0.88±0.16 | 0.95±0.03 | 0.73±0.02 | 0.72±0.02 | 0.91±0.10 | 0.90±0.11 | 0.96±0.03 | 0.94±0.03 |
| CP02 | | | | | | | | | | | |
| S01 | 0.68±0.04 | 0.56±0.03 | ----- | ----- | 0.99±0.00 | ----- | ----- | ----- | ----- | 0.96±0.05 | 0.94±0.08 |
| S02 | 0.66±0.03 | 0.55±0.05 | 0.58±0.04 | 0.59±0.03 | 0.97±0.02 | 0.68±0.01 | 0.57±0.04 | 0.67±0.02 | 0.60±0.04 | 0.94±0.10 | 0.92±0.12 |
| S03 | 0.63±0.04 | 0.58±0.02 | 0.63±0.04 | 0.68±0.04 | 0.92±0.09 | 0.62±0.01 | 0.60±0.03 | 0.64±0.04 | 0.60±0.03 | 0.92±0.13 | 0.93±0.11 |
| S04 | 0.71±0.03 | 0.69±0.03 | 0.64±0.04 | 0.71±0.03 | 0.89±0.15 | 0.69±0.02 | 0.69±0.02 | 0.73±0.01 | 0.69±0.04 | 0.87±0.16 | 0.92±0.13 |
| S05 | 0.76±0.03 | 0.70±0.04 | 0.68±0.05 | 0.74±0.04 | 0.83±0.20 | 0.71±0.04 | 0.70±0.05 | 0.78±0.02 | 0.74±0.02 | 0.88±0.13 | 0.90±0.11 |
| Avg | 0.69±0.04 | 0.61±0.07 | 0.63±0.03 | 0.68±0.06 | 0.92±0.06 | 0.68±0.04 | 0.64±0.05 | 0.70±0.05 | 0.66±0.06 | 0.91±0.04 | 0.92±0.01 |
| CP03 | | | | | | | | | | | |
| S01 | 0.62±0.05 | 0.52±0.04 | ----- | ----- | 0.98±0.01 | ----- | ----- | ----- | ----- | 0.99±0.01 | 0.96±0.05 |
| S02 | 0.65±0.04 | 0.62±0.01 | 0.56±0.04 | 0.57±0.03 | 0.99±0.01 | 0.68±0.02 | 0.61±0.03 | 0.69±0.04 | 0.62±0.03 | 0.90±0.19 | 0.95±0.06 |
| S03 | 0.69±0.02 | 0.63±0.06 | 0.70±0.06 | 0.61±0.04 | 0.98±0.02 | 0.68±0.04 | 0.67±0.03 | 0.69±0.04 | 0.68±0.03 | 0.92±0.12 | 0.90±0.13 |
| S04 | 0.69±0.01 | 0.65±0.03 | 0.60±0.05 | 0.69±0.04 | 0.90±0.16 | 0.69±0.02 | 0.65±0.06 | 0.68±0.03 | 0.67±0.07 | 0.86±0.15 | 0.93±0.08 |
| S05 | 0.69±0.01 | 0.71±0.02 | 0.70±0.03 | 0.68±0.04 | 0.82±0.21 | 0.72±0.03 | 0.69±0.03 | 0.71±0.06 | 0.72±0.02 | 0.93±0.10 | 0.92±0.09 |
| S06 | 0.66±0.03 | 0.66±0.02 | 0.54±0.01 | 0.68±0.03 | 0.87±0.15 | 0.67±0.02 | 0.66±0.02 | 0.68±0.05 | 0.68±0.04 | 0.93±0.12 | 0.91±0.10 |
| S07 | 0.69±0.02 | 0.64±0.03 | 0.72±0.02 | 0.88±0.12 | 0.80±0.16 | 0.72±0.02 | 0.64±0.03 | 0.82±0.13 | 0.77±0.08 | 0.96±0.06 | 0.91±0.11 |
| S08 | 0.67±0.02 | 0.65±0.03 | 0.73±0.02 | 0.81±0.10 | 0.82±0.14 | 0.68±0.03 | 0.68±0.02 | 0.68±0.01 | 0.77±0.12 | 0.93±0.10 | 0.90±0.11 |
| S09 | 0.60±0.08 | 0.54±0.06 | 0.65±0.03 | 0.86±0.14 | 0.76±0.15 | 0.66±0.03 | 0.67±0.04 | 0.83±0.15 | 0.86±0.13 | 0.93±0.13 | 0.89±0.13 |
| S10 | 0.59±0.04 | 0.53±0.02 | 0.72±0.09 | 0.99±0.00 | 0.77±0.14 | 0.64±0.05 | 0.72±0.02 | 0.96±0.02 | 0.96±0.02 | 0.85±0.17 | 0.88±0.14 |
| S11 | 0.65±0.04 | 0.61±0.01 | 0.63±0.04 | 0.81±0.2 | 0.64±0.08 | 0.65±0.00 | 0.64±0.02 | 0.86±0.17 | 0.82±0.17 | 0.89±0.13 | 0.93±0.10 |
| Avg | 0.65±0.03 | 0.62±0.06 | 0.65±0.07 | 0.76±0.13 | 0.85±0.10 | 0.68±0.02 | 0.66±0.03 | 0.76±0.09 | 0.76±0.10 | 0.92±0.04 | 0.92±0.02 |
| CP04 | | | | | | | | | | | |
| S01 | 0.73±0.01 | 0.62±0.02 | ----- | ----- | 0.98±0.01 | ----- | ----- | ----- | ----- | 0.99±0.00 | 0.99±0.01 |
| S02 | 0.75±0.0 | 0.70±0.03 | 0.72±0.02 | 0.72±0.02 | 0.94±0.07 | 0.75±0.02 | 0.72±0.03 | 0.75±0.03 | 0.70±0.02 | 0.97±0.04 | 1.00±0.00 |
| S03 | 0.68±0.03 | 0.50±0.01 | 0.60±0.02 | 0.56±0.03 | 0.86±0.12 | 0.70±0.04 | 0.55±0.05 | 0.73±0.02 | 0.53±0.02 | 0.91±0.10 | 0.89±0.19 |
| S05 | 0.76±0.03 | 0.75±0.01 | 0.74±0.06 | 0.79±0.02 | 0.88±0.13 | 0.75±0.03 | 0.74±0.03 | 0.76±0.03 | 0.74±0.03 | 0.94±0.08 | 0.95±0.09 |
| S06 | 0.78±0.03 | 0.61±0.02 | 0.60±0.02 | 0.64±0.03 | 0.86±0.16 | 0.75±0.03 | 0.58±0.02 | 0.80±0.04 | 0.59±0.04 | 0.95±0.07 | 0.91±0.15 |
| S07 | 0.75±0.01 | 0.76±0.01 | 0.75±0.03 | 0.90±0.08 | 0.91±0.10 | 0.76±0.01 | 0.75±0.02 | 0.96±0.06 | 0.88±0.08 | 0.93±0.0 | 0.95±0.09 |
| Avg | 0.74±0.03 | 0.66±0.09 | 0.68±0.07 | 0.72±0.17 | 0.90±0.04 | 0.74±0.02 | 0.67±0.09 | 0.80±0.08 | 0.69±0.12 | 0.95±0.03 | 0.95±0.04 |

### D. Statistical Analysis

The One-Way ANOVA showed a highly significant protocol effect on accuracy for each participant individually (all overall $p < 0.001$). Adjusted pairwise tests (Tukey HSD) indicated that Protocols 4 and 7 consistently and significantly outperformed the baseline models with limited constraints (such as Protocols 1, 2, and 5) across subjects ($p < 0.05$, often $p < 0.001$). Crucially, the difference between Protocols 4 and 7 was not statistically significant in any of the participants ($p \geq 0.05$). This suggests that extensive cross-subject pretraining effectively captures invariant MRCP patterns, leaving minimal scope for additional calibration to significantly improve performance beyond the established theoretical limit. The same relational pattern held robustly for both X = 25% and 50% splits.

### E. ROC

Figure 5 presents the ROC curves for all seven training protocols using the pooled test data from all participants. Consistent with the accuracy analysis, Protocols 4 and 7 demonstrated the strongest discriminative performance, achieving the highest AUC values of 0.952 and 0.940, respectively. Protocol 3 also demonstrated strong performance (AUC = 0.828). In contrast, Protocols 1, 2, and 5 exhibited substantially lower AUC values (0.713–0.746), indicating reduced robustness when relying on limited within-session data or previous-session calibration alone.

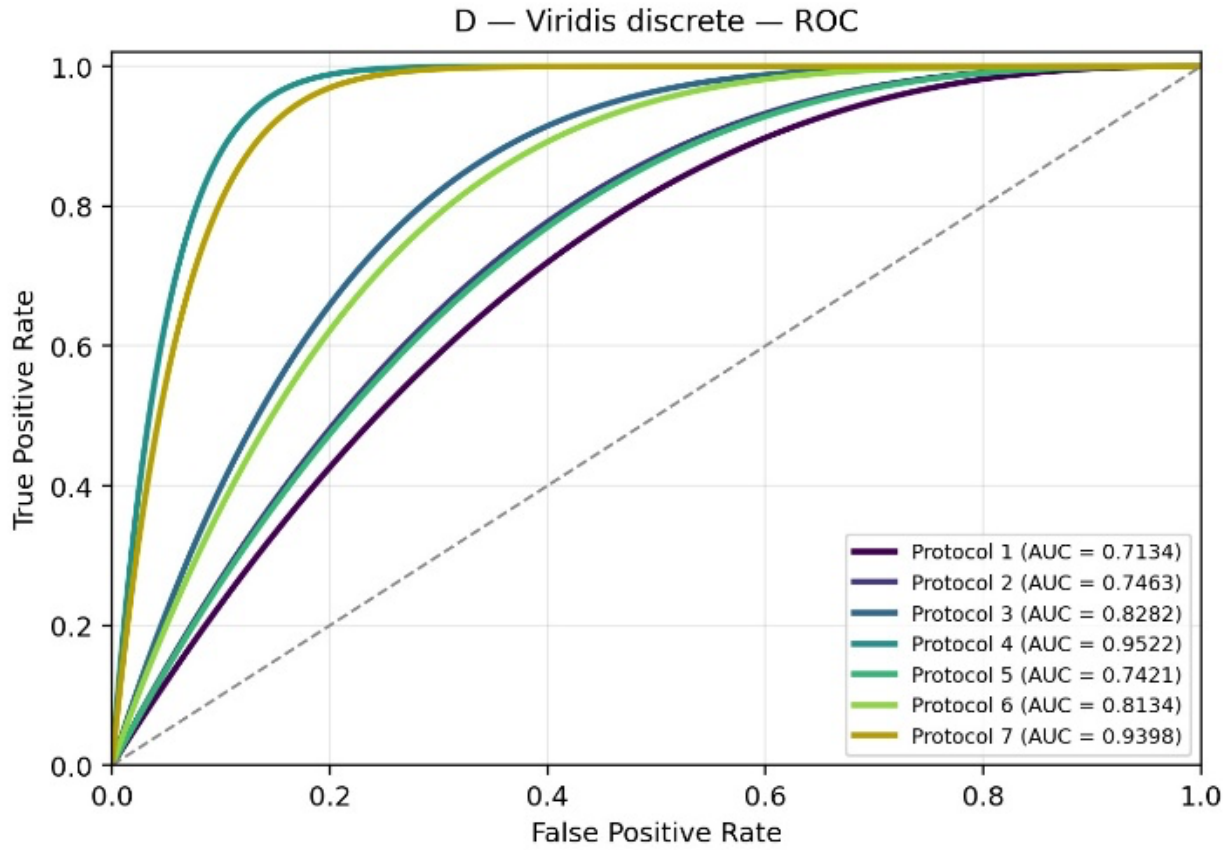

Fig. 5. Receiver operating characteristic (ROC) curves for the seven evaluated training protocols using pooled test data from all participants. The area under the curve (AUC) is reported for each protocol. The dashed diagonal line represents chance-level classification (AUC = 0.5).

## IV Discussion

The primary objective of this study was to develop a calibration-efficient deep-learning framework capable of accurately detecting MRCP-based motor intent in children with CP during a *cue-based* motor-attempt BCI paradigm, thereby addressing one of the major barriers to the clinical adoption of pediatric BCI systems. Lengthy BCI setup procedures—implemented to manage session-to-session and subject-to-subject EEG variability—pose a major barrier to clinical use and diminish user engagement. We leveraged previously collected data across seven model-training protocols to identify those that deliver high accuracy with minimal within-session calibration burden.

Our principal findings show that broad cross-subject pretraining (Protocol 4) and cross-subject pretraining coupled with brief calibration on the target session (Protocol 7) achieved the highest detection accuracy, both consistently exceeding 95%. Notably, Protocol 4 required no within-session calibration prior to evaluation, yet still yielded excellent performance. These results demonstrate that incorporating data from prior sessions and subjects can substantially reduce or even eliminate the need for in-session calibration, enabling more practical BCI-NFT systems, maximizing the time available for motor training. Both protocols benefit from aggregating a larger pool of diverse trials, which stabilizes idiosyncratic MRCP morphology and mitigates session-wise nonstationarity. Both mechanisms align with prior transfer learning research, which demonstrates that leveraging and lightly adapting pre-existing knowledge improves generalization across subjects and sessions [30], [34], [35].

It is important to note that Protocol 4 does not eliminate model updating altogether; rather, it eliminates the need for explicit within-session calibration. At initial deployment, the model is trained exclusively on data from previously enrolled participants and evaluated on the target participant without any subject-specific calibration. As new sessions are collected, the model is incrementally updated using these data, allowing personalization to occur passively through routine therapy sessions rather than through dedicated calibration procedures. Consequently, Protocol 4 may be viewed as a cumulative subject-adaptive framework that continuously updates over time while avoiding additional setup burden.

Our results demonstrate that the proposed Bi-LSTM framework can achieve high motor-intent detection accuracy in children with CP when combined with appropriate training and transfer-learning strategies. Bi-LSTM networks are particularly well suited for MRCP decoding because they capture temporal dependencies in EEG signals that evolve over hundreds of milliseconds prior to movement onset. By modeling both forward and backward temporal context, the network learns latent spatiotemporal representations of the slow, low-frequency (0.05-10 Hz) MRCP dynamics and latent features with reduced reliance on manually engineered features [37], while remaining computationally efficient for EEG applications [38]. Within the cue-based motor-attempt paradigm employed in this study, the cross-subject models (Protocols 4 and 7) exceeded 95% accuracy, substantially outperforming previously reported MI and MRCP studies, including Roy et al. (~71%) [26], Tibrewal et al. (~69%) [27], and Ingolfsson et al.'s EEG-TCNet (75–85%) [28]. These findings suggest that the combination of deep temporal modeling and cumulative cross-subject learning enables the extraction of generalized MRCP representations that are robust to the substantial EEG variability commonly observed across sessions and participants.

The high accuracies reported in the present study should be interpreted within the context of the *cue-based* BCI paradigm employed. In *cue-based* or synchronous BCIs, participants perform motor attempts in response to externally presented cues, allowing EEG analysis to be constrained to well-defined temporal windows surrounding the expected movement. Consequently, the classifier was tasked with distinguishing rest from motor-preparation periods within a controlled experimental framework referenced to verified dorsiflexion onset. This differs from self-paced or asynchronous BCIs, in which motor intentions must be continuously detected from ongoing EEG without prior knowledge of when a movement attempt will occur. While the *cue-based* design was appropriate for systematically evaluating calibration and transfer-learning strategies, future studies should determine whether the proposed framework can maintain similar performance under fully self-paced neurofeedback conditions.

The robustness of Protocols 4 and 7 was further supported by both F1-score and ROC analyses. These protocols consistently achieved the highest F1-scores, indicating a balanced tradeoff between precision and recall, which is particularly important in associative-learning BCIs where false-positive detections may weaken the temporal coupling between motor intent and sensory feedback. Similarly, AUC values exceeding 0.93 demonstrated excellent separability between motor-intent and rest EEG states. Together, the high F1-scores and AUC values indicate that the observed performance was not driven by a favorable classification threshold alone, but rather by stable and generalized MRCP representations that remained robust despite substantial inter-session and inter-subject variability. Overall, these findings further support the effectiveness of cross-subject pretraining and incremental adaptation for calibration-efficient MRCP detection in children with CP.

When compared with previously reported transfer learning strategies, our Protocol 7—cross-subject pretraining followed by brief, session-specific fine-tuning—achieved markedly higher accuracy without requiring complex adversarial distribution alignment modules. Hybrid domain-adaptation with few-shot fine-tuning (few calibration trials) approaches, such as DFF-Net (Lu et al.) [33], combine distribution-alignment modules with small, labeled datasets to improve cross-subject generalization, yet their accuracies remained in the 80–85% range. Other transfer-learning methods in the MI literature, including task-free transfer learning [31], cross-task transfer from bilateral to unilateral movement-intention decoding [35], and cross-subject CNN fusion architectures by Chowdhury et al. [32], report accuracies in the 75–88% range.

Compared with single-session or baseline calibration strategies, Protocol 4 demonstrates the ability to operate without any calibration for new sessions once a sufficient amount of data has been accumulated across source subjects. In this protocol, model training is initialized using data from all previous participants; therefore, at Session 1 (S01 in Table IV) the model has not yet been exposed to data from the target subject. Despite this, detection accuracy remains stable across subjects, as evidenced by the virtually unchanged near-perfect performance beginning at S01. This indicates that Protocol 4 can be deployed immediately without subject-specific calibration once adequate longitudinal data has been collected from previous subjects.

In contrast to prior approaches that rely on a dedicated calibration phase—such as Biasiucci et al. [1], who used an extended pretraining calibration session to train an MRCP detector reused throughout the intervention, or Rao et al. [24], who required an initial extended calibration phase before implementing a once-calibration strategy across sessions—Protocol 4 instead exploits cumulative, data-driven aggregation of cross-subject trials collected during prior studies. By progressively incorporating these longitudinal data, the model adapts naturally to session-to-session EEG variability, effectively eliminating the need for explicit within-session calibration after a critical number of source trials have been reached.

The present findings directly address several limitations previously identified in MRCP-based pediatric BCI systems. Jochumsen et al. reported that motor-intent detection performance of a *cue-based* BCI paradigm, in individuals with CP, declined from approximately 75% during same-day calibration to ~60% when models were transferred across sessions and further to ~40–53% in cross-subject calibration conditions, highlighting the difficulty of maintaining stable decoding under pediatric EEG variability [25]. In contrast, Protocols 4 and 7 in the present study maintained consistently high performance (>90–95% accuracy; AUC > 0.93) despite relying on minimal—or, in the case of Protocol 4, no within-session subject-specific calibration.

From a clinical perspective, reducing calibration burden is particularly important in pediatric neurorehabilitation, where attention span, fatigue, and session duration can directly influence treatment effectiveness. By minimizing or eliminating dedicated calibration procedures, Protocols 4 and 7 increase the proportion of each therapy session available for active neurofeedback training. This may improve user engagement, reduce frustration associated with prolonged setup procedures, and facilitate the integration of MRCP-based BCI-NFT into routine clinical practice.

This study has several limitations. First, although the dataset comprised 2,637 trials collected across 27 sessions, only four participants were included, limiting the ability to fully characterize inter-subject variability within the heterogeneous CP population. Second, motor-intent labels were derived from predefined peri-movement time windows referenced to verified dorsiflexion onset rather than from dynamically detected MRCP events, which may introduce some labeling uncertainty. Third, the present evaluation was performed offline within a cue-based paradigm, where the timing of motor attempts was constrained to predefined windows. Consequently, the reported performance may not fully reflect the challenges associated with continuous, self-paced BCI operation.

**Future work** will benchmark the proposed framework against established EEG deep-learning architectures, such as EEGNet [39], as well as other recently proposed models [31], and evaluate performance on publicly available datasets using identical preprocessing and labeling procedures. Additional studies will investigate low-effort calibration strategies, including adaptation using resting-state EEG, and assess whether the observed benefits of cumulative learning and transfer learning translate to improved decoding performance in real-time applications. Finally, prospective clinical studies are needed to determine whether the calibration-efficient protocols developed here improve intervention efficiency, user engagement, adherence, training dose, and ultimately functional outcomes during pediatric BCI-based neurofeedback rehabilitation.

## V. Conclusion

This study demonstrates that accurate, calibration-efficient detection of motor intent from MRCP is feasible in children with CP using deep learning and transfer learning. By systematically comparing seven training strategies, we show that both zero-calibration cross-subject pretraining and cross-subject pretraining with brief calibration can achieve robust decoding accuracy consistently exceeding 95%, substantially outperforming conventional session-specific and session-to-session approaches. Notably, generalized cross-subject training eliminated the need for any in-session calibration after sufficient prior data were collected, addressing one of the most persistent barriers to real-world BCI-NFT deployment.

The strong performance of the proposed Bi-LSTM framework suggests that explicitly modeling the temporal evolution of MRCPs is advantageous for motor-intent detection in children with CP. Unlike oscillatory EEG features that are primarily characterized by changes in spectral power, MRCPs manifest as slow, gradually evolving cortical potentials that unfold over hundreds of milliseconds prior to movement onset. By leveraging recurrent memory mechanisms and attention-based temporal weighting, the proposed framework can capture these long-range temporal dependencies and emphasize the most informative portions of the premovement EEG signal. This capability may contribute to the robustness of the proposed approach across sessions and subjects despite the substantial variability commonly observed in pediatric EEG recordings.

By reducing setup time while preserving reliable motor-intent detection, the proposed framework improves clinical feasibility and paves the way for scalable, real-time neurorehabilitation interventions in children with CP.

## Acknowledgment

This research was supported in part by the Intramural Research Program of the National Institutes of Health (NIH). The contributions of the NIH author(s) are considered Works of the United 686 States Government. The findings and conclusions presented in this paper are those of the author(s) and do not necessarily reflect the views of the NIH or the U.S. Department of Health and Human Services.